\documentclass[12pt,a4paper]{article} 
\usepackage{amsmath}
\usepackage{amssymb}
\usepackage{latexsym}
\usepackage{amsthm}
\usepackage{amsbsy}
\usepackage{amstext}
\usepackage[dvips]{graphicx}

\def\lsim{\mathrel{\lower2.5pt\vbox{\lineskip=0pt\baselineskip=0pt 
           \hbox{$<$}\hbox{$\sim$}}}} 
\def\gsim{\mathrel{\lower2.5pt\vbox{\lineskip=0pt\baselineskip=0pt 
           \hbox{$>$}\hbox{$\sim$}}}}

\def\p{\partial}

\def\dis{\displaystyle}

\begin{document} 
\begin{flushright}
AUE-06-01\\ hep-th/0611075
\end{flushright}

\vspace{10mm}

\begin{center}
{\Large \bf 
Cosmological solutions for a model with a $1/H^{2}$ term}

\vspace{20mm}
 Masato Ito 
 \footnote{mito@auecc.aichi-edu.ac.jp}
\end{center}

\begin{center}
{
{}Department of Physics, Aichi University of Education, Kariya, 
448-8542, JAPAN
}
\end{center}

\vspace{25mm}

\begin{abstract}
We drive the cosmological solutions of five-dimensional model with $1/H^{2}$
term $(H^{2}\equiv H_{MNPQ}H^{MNPQ})$, where $H_{MNPQ}$ is 4-form
field strength.
The behaviors of the scale factors and the scalar potential in effective
theory are examined.
As a consequence, we show that the universe changes from decelerated
expansion to accelerated expansion in Einstein frame of the
four-dimensional theory.
\end{abstract}
\newpage 
\baselineskip=6.0mm

Observational evidences indicated that dark energy which may generate
the current cosmic acceleration dominates about
70 percent of the critical energy density
\cite{Riess:1998cb,Perlmutter:1998np}.
WMAP supports strongly the fact that the dark energy exists in present time
\cite{Lewis:2002ah,Bennett:2003bz}.
In cosmology and particle physics, it is an important problem what the
origin for driving cosmic acceleration is.
Higher dimensional theories, such as string/M-theory and braneworlds, include
scalar fields or anti-symmetric tensor fields which play an important role in
cosmological models.
In this paper, we focus on a particular model with an inverse squared
field strength and investigate the cosmological evolution of the model.

We consider the five-dimensional model with $1/H^{2}$ term,
where $H^{2}\equiv H_{MNPQ}H^{MNPQ}$ and $H_{MNPQ}$ denotes 4-form
field strength for 3-form anti-symmetric tensor field.
The action is given by
\begin{eqnarray}
S=\int\; d^{4}xdy\sqrt{-g}\left(\frac{1}{2}{\cal R}+2\cdot 4!\;\frac{1}{H^{2}}\right)
\,,\label{eqn1}
\end{eqnarray}
where $y$ means the fifth dimension, the fundamental scale is set to
unity and $g$ is determinant of five dimensional metric.
The introduction of the $1/H^{2}$ term is motivated by the particular
model of \cite{Kim:2000mc}.
The model indicated that the $1/H^{2}$ term is indispensable
when solving the cosmological constant problem without fine-tuning
between parameters in the Lagrangian.
As mentioned in \cite{Kim:2000mc,Kim:2001ez}, the unusual $1/H^{2}$ term
makes sense only if $H^{2}$ must develop a vacuum expectation value on the
order of the fundamental scale. 
We expect that the term may be generated by dynamics of quantum
gravity.
If the VEV of $1/H^{2}$ is significant effect in the history of universe, it is
natural to think that the term affects cosmic evolution.  
Thus, we have an interest in the evolution of universe
in the present model.

We assume that the form of the metric is given by
\begin{eqnarray}
ds^{2}=-dt^{2}+a^{2}(t)d\vec{x}^{2}+b^{2}(t)dy^{2}\,,\label{eqn2}
\end{eqnarray}
where $a(t)$ is the scale factor of three-dimensional space and $b(t)$
is the scale factor of an extra dimension.
The compact space of the extra dimension is assumed to be maximally
symmetric space such as $S^{1}$. 
This metic corresponds to higher dimensional Kasner metric with
asymmetric scale factors \cite{Arkani-Hamed:1999gq}.

Variation with respect to the metric leads to Einstein equation
\begin{eqnarray}
{\cal R}_{MN}-\frac{1}{2}g_{MN}{\cal R}
=2\cdot 4!\left(
\frac{8}{H^{4}}H_{MPQR}H^{PQR}_{N}+g_{MN}\frac{1}{H^{2}}
\right),\label{eqn3}
\end{eqnarray}
where $M,N=0,1,2,3,5$ denote the five-dimensional index.
The field equation for 3-form antisymmetric tensor $A_{PQR}$ is
\begin{eqnarray}
\p_{M}\left(
\frac{\sqrt{-g}H^{MPQR}}{H^{4}}
\right)=0\,.\label{eqn4}
\end{eqnarray}
Referring to the Freund-Rubin ansatz \cite{Kim:2000mc,Freund:1980xh},
it is assumed that 4-form field strength is
\begin{eqnarray}
H_{\mu\nu\rho\sigma}\propto \sqrt{-g}\;b^{-2/3}\;\epsilon_{\mu\nu\rho\sigma}\,,
\label{eqn5}
\end{eqnarray}
where $\mu,\nu,\cdots=0,1,2,3$ denote indices of the four dimensions. 
Since Eq.(\ref{eqn5}) leads to $\dis \sqrt{-g}H^{\mu\nu\rho\sigma}/H^{4}=$const,
Eq.(\ref{eqn4}) is satisfied. 

The $(00),(ij),(55)$ components of Einstein equation are 
\begin{align}
3\left(\frac{\dot{a}}{a}\right)^{2}+3\frac{\dot{a}}{a}\frac{\dot{b}}{b}&=
\frac{6}{\alpha}b^{-\frac{2}{3}},\label{eqn6}\\
2\frac{\ddot{a}}{a}+\frac{\ddot{b}}{b}+
\left(\frac{\dot{a}}{a}\right)^{2}+2\frac{\dot{a}}{a}\frac{\dot{b}}{b}
&=\frac{6}{\alpha}b^{-\frac{2}{3}},\label{eqn7}\\
3\frac{\ddot{a}}{a}+3\left(\frac{\dot{a}}{a}\right)^{2}&=
\frac{2}{\alpha}b^{-\frac{2}{3}},\label{eqn8}
\end{align}
where the dot means the derivative with respect to $t$ and $\alpha$ is a
positive constant originating from the proportionality constant in Eq.(\ref{eqn5}).
Since it is difficult to solve Eqs.(\ref{eqn6})-(\ref{eqn8}), 
we define a new variable $\tau$ as 
\begin{eqnarray}
d\tau=a^{-3}b^{-1}dt\,.\label{eqn9}
\end{eqnarray}
Solving the differential equations as shown in Appendix, the scale
factors can be expressed in terms of $\tau$
\begin{align}
a(\tau)&=\left[
\sinh^{2}\left(\frac{2}{3}|C|\tau\right)
\right]^{-\frac{3}{28}}
\exp\left(-\frac{1}{7}C\tau\right)\,,\label{eqn10}\\
b(\tau)&=\left[
\sinh^{2}\left(\frac{2}{3}|C|\tau\right)
\right]^{-\frac{15}{56}}
\exp\left(\frac{9}{14}C\tau\right)\,,\label{eqn11}
\end{align}
where $C$ is a integration constant.

We would like to consider the universe of this model in the Einstein
frame.
In order to transform the action with metric $(\ref{eqn2})$ to Einstein
frame, the conformal transformation is applied as follows
\begin{eqnarray}
\bar{g}^{(4)}_{\mu\nu}=bg^{(4)}_{\mu\nu}\,.\label{eqn12}
\end{eqnarray}
That is, the effective Lagrangian\footnote{Since the action is $
\dis S=\int d^{4}xdy\sqrt{-g^{(5)}}\;L=\int d^{4}xdy\; L_{\rm eff}$, we can obtain
(\ref{eqn13}) by substituting (\ref{eqn12}) into $L_{\rm eff}$.} is given by
\begin{eqnarray}
L_{\rm eff}=
\sqrt{-\bar g^{(4)}}\left(
\frac{1}{2}\bar{\cal R}^{(4)}-3b^{-1}\bar{g}^{(4)\mu\nu}\p_{\mu}b^{1/2}\p_{\nu}b^{1/2}
+2\;4!\;b^{-1}\frac{1}{H^{2}}
\right)\,,\label{eqn13}
\end{eqnarray}
where the barred quantities represent ones in the Einstein frame.
Thus the scale factor $b$ behaves as a scalar field in the
four-dimensional theory.
Moreover, according to the assumption of an compact extra dimension with
maximally symmetric space, the contributions of the curvature of the
extra dimension are vanishing in $L_{\rm eff}$.

By introducing the scalar field $\Phi$ as 
\begin{eqnarray}
b=e^{\sqrt{2/3}\;\Phi}\,,\label{eqn14}
\end{eqnarray}
we can obtain the Lagrangian with the canonically kinetic term
\begin{eqnarray}
L_{\rm eff}=
\sqrt{-\bar g^{(4)}}\left(
\frac{1}{2}\bar{\cal R}^{(4)}
-\frac{1}{2}\bar{g}^{(4)\mu\nu}\p_{\mu}\Phi\p_{\nu}\Phi
-\frac{2}{\alpha}e^{-\sqrt{50/27}\;\Phi}
\right)\,,\label{eqn15}
\end{eqnarray}
where we used $(\ref{eqn5})$.
The effective scalar potential for a scalar field $\Phi$ is
\begin{eqnarray}
V_{\rm eff}=\frac{2}{\alpha}e^{-\sqrt{50/27}\;\Phi}\,.\label{eqn16}
\end{eqnarray}
The exponent of scalar potential plays an important role of the
evolution of universe.
We discuss it later.

In order to investigate cosmological evolution of the present
cosmological model, we examine the sign of the derivative with respect
to the time of the scale factors for four-dimensional observer.
From Eq.(\ref{eqn12}), the time and the scale factor in
Einstein frame are given by
\begin{eqnarray}
d\bar{t}=b^{1/2}dt\;,\;\bar{a}=b^{1/2}a\,.\label{eqn17}
\end{eqnarray}
Before examining the behavior of $\bar{a}$, it is necessary to see
the relation between $\bar{t}$ and $\tau$.
From Eqs.(\ref{eqn9}) and (\ref{eqn17}), one gets
\begin{eqnarray}
d\bar{t}=a^{3}b^{3/2}d\tau=
\left[
\sinh^{2}\left(\frac{2}{3}|C|\tau\right)
\right]^{-\frac{81}{112}}
\exp\left(\frac{15}{28}C\tau\right)
d\tau\,.\label{eqn18}
\end{eqnarray}
We can explore the asymptotic relation for the limit 
$|\tau|\to 0$ and $|\tau|\to \infty$.

For $|\tau|\to 0$, according to Eq.(\ref{eqn18}),
$\tau\to +0$ maps to $\bar{t}\to -\infty$ via 
$\bar{t}\sim -\tau^{-25/56}$ and
$\tau\to -0$ maps to $\bar{t}\to +\infty$ via 
$\bar{t}\sim (-\tau)^{-25/56}$.

For $|\tau|\to \infty$, we must carefully handle Eq.(\ref{eqn18}).
In the case of $C>0$, $\tau\to +\infty$ maps to $\bar{t}\to -0$ via 
$\bar{t}\sim -\exp(-3C\tau/7)$ and 
$\tau\to -\infty$ maps to $\bar{t}\to +0$ via 
$\bar{t}\sim \exp(3C\tau/2)$.
On the other hands,
in the case of $C<0$, $\tau\to +\infty$ maps to $\bar{t}\to -0$ via 
$\bar{t}\sim -\exp(-3|C|\tau/2)$ and 
$\tau\to -\infty$ maps to $\bar{t}\to +0$ via 
$\bar{t}\sim \exp(3|C|\tau/7)$.
Although the dependence of $\tau$ is different, regardless of the sign of
the integration constant $C$, it turns out that the sign and direction of
$\bar{t}$ and $\tau$ are reverse.

From Eqs.(\ref{eqn9}) and (\ref{eqn17}), the derivatives of $\bar{a}$
are given by
\begin{align}
\frac{d}{d\bar{t}}\bar{a}&=
 a^{-3}b^{-3/2}\left(H_{a}+\frac{1}{2}H_{b}\right)\,,\label{eqn19}\\
\frac{d^{2}}{d\bar{t}^{2}}\bar{a}&=
a^{-6}b^{-3}\left\{
-2\left(H_{a}+\frac{1}{2}H_{b}\right)^{2}
+H^{\prime}_{a}+\frac{1}{2}H^{\prime}_{b}
\right\}\,,\label{eqn20}
\end{align}
where $H_{a}=a^{\prime}/a$ and $H_{b}=b^{\prime}/b$, the prime denotes
the derivative of $\tau$.

Following Eqs.(\ref{eqn19}) and (\ref{eqn20}),
the relevant terms which determine the signs of $d\bar{a}/d\bar{t}$ and 
$d^{2}\bar{a}/d\bar{t}^{2}$ can be expressed as
\begin{align}
{\rm sgn}\left(\frac{d}{d\bar{t}}\bar{a}\right)&= {\rm sgn}\left(
-9\coth\frac{2}{3}|C|\tau+5\frac{C}{|C|}\right)\,,
\label{eqn21}\\
{\rm sgn}\left(\frac{d^{2}}{d\bar{t}^{2}}\bar{a}\right)&=
{\rm sgn}\left(
3\coth^{2}\frac{2}{3}|C|\tau
+90\frac{C}{|C|}\coth\frac{2}{3}|C|\tau-109
\right)\,,\label{eqn22}
\end{align}
where ${\rm sgn}(x)$ denotes the sign function.
The signs of $d\bar{a}/d\bar{t}$ and
$d^{2}\bar{a}/d\bar{t}^{2}$ are explicitly shown below.
\begin{center}
\begin{tabular}{c|ccccccccc}
$\tau$ & $-\infty$ &  & $\tau_{1}$ & & $0$&  & $\tau_{2}$ & & $+\infty$\\ \hline
$d\bar{a}/d\bar{t}$ & &$+$ & & & $0$ & && $-$&\\
$d^{2}\bar{a}/d\bar{t}^{2}$ & & $-$ & $0$ & & $+$ & &$0$ & $-$ &
\end{tabular}
\end{center}
According to Eq.(\ref{eqn22}),
the values of $\tau_{1}$ and $\tau_{2}$ depend on the sign of $C$.
For $C>0$, we have $\tau_{1}=-(3/4C)\log 2(2+\sqrt{3})/7$ and 
$\tau_{2}=(3/4C)\log 7(2+\sqrt{3})/2$.
For $C<0$, we have $\tau_{1}=-(3/4|C|)\log 7(2+\sqrt{3})/2$ and 
$\tau_{2}=(3/4|C|)\log 2(2+\sqrt{3})/7$.

The survey of the behaviors of $\bar{a}$ for $\tau$ is
illustrated in Fig.1.
Here $d\bar{a}/d\bar{t}>0$ and $d\bar{a}/d\bar{t}<0$ correspond to
expansion and contraction, respectively.
On the other hands, $d^{2}\bar{a}/d\bar{t}^{2}>0$ and
$d^{2}\bar{a}/d\bar{t}^{2}<0$ correspond to
acceleration and deceleration, respectively.

\begin{figure}
\begin{center}
\includegraphics[width=13cm]{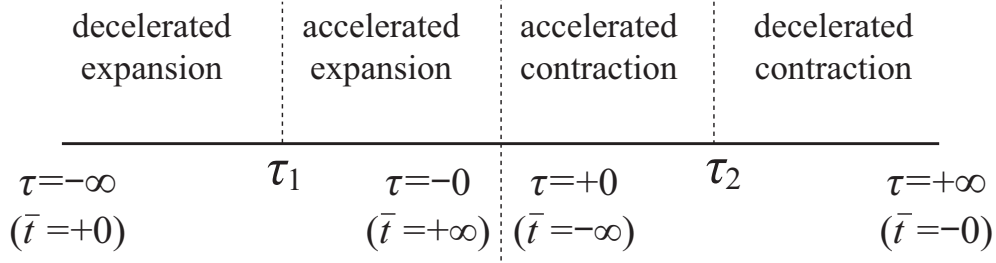}
\caption{The cosmological evolution for
 $\tau\;(\bar{t})$ is shown. The horizontal line is $\tau\;(\bar{t})$.}
\label{fig1}
\end{center}
\end{figure}
From the above table, the expansion occurs during
$-\infty<\tau<-0\;(+0<\bar{t}<+\infty)$ and the contraction does during
$+0<\tau<+\infty\;(-\infty<\bar{t}<-0)$.
In addition, we can see that the periods of
$\tau_{1}<\tau<-0$ and $+0<\tau<\tau_{2}$ are accelerated phase,
the periods of $-\infty<\tau<\tau_{1}$ and $\tau_{2}<\tau<+\infty$ are
decelerated phase.
As shown in the Fig.1, the era of accelerated expansion exists in the range
$\bar{t}(\tau=\tau_{1})<\bar{t}<+\infty$.
Interestingly, it turns out that the universe changes from the
decelerated expansion to accelerated expansion during 
$+0<\bar{t}<+\infty(-\infty<\tau<-0)$.

Although we cannot obtain the explicit form of scale factor
$\bar{a}(\bar{t})$, the asymptotic behavior of $\bar{a}$ can be
evaluated.
We consider the range of $+0<\bar{t}<+\infty(-\infty<\tau<-0)$, since
$\bar{t}$ is the proper time of the Einstein frame for the four dimensional
observer. 

For $\bar{t}\to +0 (\tau\to -\infty)$, Eqs.(\ref{eqn10}) and
(\ref{eqn18}) lead to
\begin{eqnarray}
\bar{a}\sim \bar{t}\;^{\frac{1}{3}}\,.\label{eqn23}
\end{eqnarray}
At initial time, $\bar{a}$ evolves through the equation of state
corresponding to $\omega=1$.
In general case, the scale factor is given by $a\sim t^{2/3(\omega+1)}$
for the equation of state $p=\omega\rho$, where $p$, $\rho$ are the pressure and
the energy density.

For $\bar{t}\to +\infty (\tau\to -0)$, we obtain
\begin{eqnarray}
\bar{a}\sim \bar{t}\;^{\frac{27}{25}}\,.\label{eqn24}
\end{eqnarray}
For sufficient late time, $\bar{a}$ evolves through the equation of state
corresponding to $\omega=-31/81$.
This era is an accelerating phase due to $-31/81<-1/3$.
Thus it turns out that the transition from $\omega=1$ (stiff) to
$\omega=-31/81$ (Quintessence) corresponds to 
from decelerated expansion to accelerated expansion.

By analysing the properties of the effective scalar potential in
Eq.(\ref{eqn16}), we confirm the
results of Eqs.(\ref{eqn23}) and (\ref{eqn24}).
We consider the evolution of the scalar field in present model.
The scalar field $\Phi$ and the velocity $d\Phi/d\bar{t}$ are given by
\begin{align}
\Phi&=\sqrt{\frac{3}{2}}
\left(
-\frac{15}{28}\log\left|\sinh\frac{2|C|}{3}\tau\right|+\frac{9}{14}C\tau
\right)\,,\label{eqn25}\\
\frac{d\Phi}{d\bar{t}}&=\frac{|C|}{14}
\left(\sinh^{2}\frac{2|C|}{3}\tau\right)^{\frac{81}{112}}e^{-\frac{15}{28}C\tau}
\left(-5\coth\frac{2|C|}{3}\tau +9\frac{C}{|C|}\right)\,.\label{eqn26}
\end{align}
The asymptotic behaviors of the scalar field and the velocity are examined.

At initial time $\bar{t}\to +0(\tau\to -\infty)$, the start point of
$\Phi$ depends on the sign of the integration constant $C$
\begin{eqnarray}
\left\{
\begin{array}{ll}
\dis C>0 & \Phi\sim C\tau \to -\infty\\ 
&\\
\dis C<0 & \dis \Phi\sim -\frac{2}{17}|C|\tau \to +\infty
\end{array}
\right.\,.\label{eqn27}
\end{eqnarray}
Furthermore, for the same limit $\tau\to -\infty$, the asymptotic
velocities are
\begin{eqnarray}
\left\{
\begin{array}{ll}
\dis C>0 & \dis\frac{d\Phi}{d\bar{t}}
\sim +e^{-43C\tau/28} \to +\infty\\ 
&\\
\dis C<0 & \dis \frac{d\Phi}{d\bar{t}}
\sim -e^{-12|C|\tau/7} \to -\infty
\end{array}
\right.\,.\label{eqn28}
\end{eqnarray}
Since the kinetic energy $|d\Phi/d\bar{t}|^{2}$ dominate over potential
energy $V(\Phi)$, we obtain $\omega=1$ because of $p=\rho$.
This is consistent with the result of Eq.(\ref{eqn23}).

At the sufficient late time $\bar{t}\to +\infty(\tau\to -0)$, from
Eqs.(\ref{eqn25}) and (\ref{eqn26}), we get
\begin{eqnarray}
\Phi\to +\infty\;,\;
\frac{d\Phi}{d\bar{t}}\to +0\,,\label{eqn29}
\end{eqnarray} 
regardless of the sign of $C$.
Since the velocity is very small,
this implies that the kinetic energy and potential energy are compatible.
In this case, the value of $\omega$ is determined by the exponent of the 
exponential potential.
In general case, the potential of $e^{-\lambda\psi}$($\psi$ is
canonically normalized field) has $\omega=\lambda^{2}/3-1$.
According to Eq.(\ref{eqn16}),
in the case of the present model with $\lambda=\sqrt{50/27}$, we obtain
\begin{eqnarray}
\omega=-\frac{31}{81}\,.\label{eqn30}
\end{eqnarray}
Obviously, this result is consistent with the behavior of scale factor of
Eq.(\ref{eqn23}).
\begin{figure}
\begin{center}
\includegraphics[width=14cm]{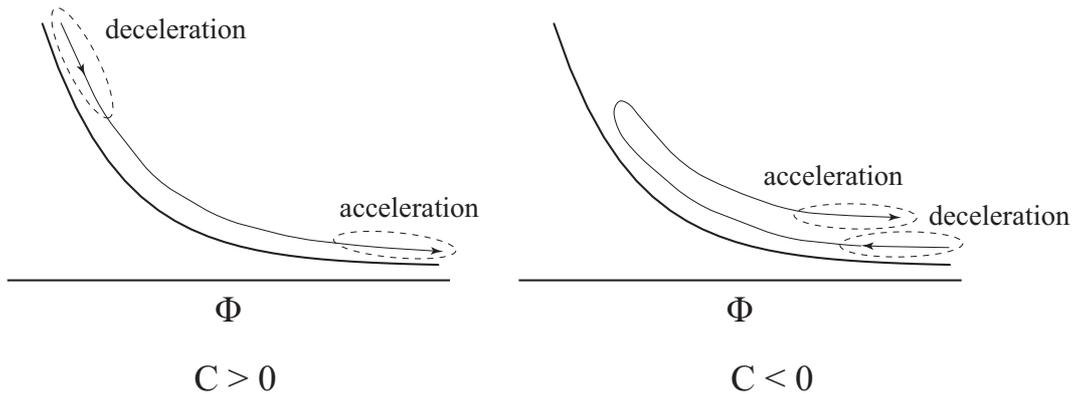}
\caption{The illustrations represent the evolution of scalar field $\Phi$.
Depending on the sign of $C$, the initial start point of $\Phi$ is
different.
For $C>0$, starting from $\Phi\to -\infty$, the scalar field goes to
$+\infty$ and rolls down hill.
For $C<0$, starting from $\Phi\to +\infty$, the scalar field climbs up 
the hill and turns around, and are rolling down hill.
}
\label{fig2}
\end{center}
\end{figure}

In Fig.2, the evolutions of a scalar field $\Phi$ are shown.
Eq.(\ref{eqn27}) indicates that the initial start point of $\Phi$ is
different for two cases.
By combining with Eqs.(\ref{eqn27})-(\ref{eqn29}), we can understand the
evolutions of $\Phi$.
For $C>0$, starting from $\Phi\to -\infty$, the scalar field rolls down
toward $\Phi\to +\infty$.
For $C<0$, starting from $\Phi\to +\infty$, the scalar field climbs up 
the hill and turns around, it rolls down toward $\Phi\to +\infty$.
There exists the transition from decelerated expansion to accelerated
expansion for each case.

In conclusion, we derived the cosmological solutions of five dimensional
model with $1/H^{2}$ term.
It was pointed out that in Einstein frame the universe changes
from decelerated expansion to accelerated expansion.
From the analysis of asymptotic behaviors of the scale factors, it was
shown that the transition is from the phase of $\omega=1$ to the phase
of $\omega=-31/81$.
Furthermore, we analysed the properties of effective scalar potential
and a scalar field in effective Lagrangian.
As a consequence, we obtained the same results about the evolutions of
the universe.
Thus we constructed a toy model in accordance with the current cosmic
acceleration by introducing an inverse squared field strength
term.
We expect that the term can be generated from the dynamics of the
underlying theories.
The present model don't have the inflationary period in early
universe.
The model should be improved, for example, the extension to more higher
dimensions is possible.
It may be necessary to consider the drastic models
in order to explain recent astronomical observation in the framework of
cosmology and particle physics.
When proposing the model of the cosmic acceleration, the present model with
$1/H^{2}$ term may have several new possibilities.

\medskip

\noindent {\bf Appendix}:

From Eqs.(\ref{eqn6})-(\ref{eqn8}),
we derive Eqs.(\ref{eqn10}) and (\ref{eqn11}) \cite{Arkani-Hamed:1999gq}.
The following transformations are performed.
\begin{eqnarray}
a=e^{X}\;,\;b=e^{Y}\,.\label{eqna1}
\end{eqnarray}
Substituting Eq.(\ref{eqna1}) into Eqs.(\ref{eqn6})-(\ref{eqn8}), the
suitable linear combinations yield the following equations
\begin{align}
\ddot{X}+\dot{X}\left(3\dot{X}+\dot{Y}\right)&=\frac{8}{3\alpha}e^{-\frac{2}{3}Y}
\label{eqna2}\,,\\
\ddot{Y}+\dot{Y}\left(3\dot{X}+\dot{Y}\right)&=\frac{20}{3\alpha}e^{-\frac{2}{3}Y}
\label{eqna3}\,,\\
\dot{X}^{2}+\dot{X}\dot{Y}&=\frac{2}{\alpha}e^{-\frac{2}{3}Y}\,.
\label{eqna4}
\end{align}
Using Eq.(\ref{eqn9}), we can rewrite
Eqs.(\ref{eqna2})-(\ref{eqna4}) as follows
\begin{align}
X^{\prime\prime}&=\frac{8}{3\alpha}e^{Z}\,,\label{eqna5}\\
Y^{\prime\prime}&=\frac{20}{3\alpha}e^{Z}\,,\label{eqna6}\\
X^{\prime 2}+X^{\prime}Y^{\prime}&=\frac{2}{\alpha}e^{Z}
\,,\label{eqna7}\\
Z&=6X+\frac{4}{3}Y\,,\label{eqna8}
\end{align}
where the prime means the derivative with respect to $\tau$.
Using Eqs.(\ref{eqna5}), (\ref{eqna6}) and (\ref{eqna8}), we have
\begin{eqnarray}
Z^{\prime\prime}=\frac{224}{9\alpha}e^{Z}\;\to\;
Z^{\prime 2}=\frac{448}{9\alpha}e^{Z}+C_{1}\,,\label{eqna9}
\end{eqnarray}  
where $C_{1}$ is a constant.
Furthermore, Eqs.(\ref{eqna5}) and (\ref{eqna6}) yield
\begin{eqnarray}
Y=\frac{5}{2}X+C\tau+C_{2}\,,\label{eqna10}
\end{eqnarray}
where $C,C_{2}$ are constants.
From Eqs.(\ref{eqna9}) and (\ref{eqna10}), Eq.(\ref{eqna7}) leads to
\begin{eqnarray}
C_{1}=\frac{16}{9}C^{2}\,.\label{eqna11}
\end{eqnarray}
Substituting Eq.(\ref{eqna11}) into Eq.(\ref{eqna9}), we obtain
\begin{eqnarray}
Z^{\prime 2}=\frac{448}{9\alpha}e^{Z}+\frac{16}{9}C^{2}\,.
\label{eqna12}
\end{eqnarray}
Solving it, we have
\begin{eqnarray}
e^{Z}=\frac{\alpha C^{2}}{28}\frac{1}{\dis\sinh^{2}\left(\frac{2}{3}|C|\tau\right)}
\,.\label{eqna13}
\end{eqnarray}
Using Eqs.(\ref{eqna1}), (\ref{eqna8}), (\ref{eqna10}) and Eq.(\ref{eqna13}),
we can obtain Eqs.(\ref{eqn10}) and (\ref{eqn11}).
Note that the absolute in argument of Eq.(\ref{eqna13}) stems from 
$\sqrt{C^{2}}=|C|$ when solving Eq.(\ref{eqna12}), $|C|$ cannot be
removed by rescaling.
Because, over the range $-\infty<\tau<+\infty$, the behaviors of the
scale factors depends on the sign of $C$ through Eq.(\ref{eqna10}).
Consequently, as shown in Fig.2, the sign of $C$ controlls the revolutions of
the scalar field.
The situations are similar to the model of Ref.\cite{Arkani-Hamed:1999gq}.

\medskip

\noindent {\bf Acknowledgements}: 
This work is supported in part by the Grants-in-Aid for Scientific
Research of the Ministry of Education, Culture, Sports, Science and
Technology of Japan (No.17740141).


\end{document}